\documentclass[aps,twocolumn,pra,10pt,showpacs,superscriptaddress]{revtex4-1}

\usepackage[utf8]{inputenc}
\usepackage{graphicx}
\usepackage{amsfonts}
\usepackage{amssymb,amsmath}
\usepackage{verbatim}
\usepackage{bbold}

\newcommand{\ket}[1]{\left| #1 \right\rangle}
\newcommand{\bra}[1]{\left\langle #1 \right|}

\newcommand{\tr}{\mathop{\mathrm{tr}}}

\newcommand{\void}[1]{}

\begin{document}

\title{Qubit interference at avoided crossings: The role of driving shape and bath coupling}

\author{Ralf Blattmann}
\author{Peter H{\"a}nggi}
\affiliation{Institut f{\"u}r Physik, Universit{\"a}t Augsburg,
Universit{\"a}tsstra{\ss}e 1, D-86153 Augsburg, Germany}

\author{Sigmund Kohler}
\affiliation{Instituto de Ciencia Materiales de Madrid, CSIC, Cantoblanco, E-28049 Madrid, Spain}

\begin{abstract}
We derive the structure of the Landau-Zener-St\"uckelberg-Majorana (LZSM)
interference pattern for a qubit that experiences quantum dissipation and
is additionally subjected to time-periodic but otherwise general driving.
A spin-boson Hamiltonian serves as model which we treat with a
Bloch-Redfield master equation in Floquet basis.  It predicts a peak
structure that depends sensitively on the operator through which the qubit
couples to the bath.  The Fourier transforms of the LZSM patterns exhibit
arc structures which reflect the shape of the driving.  These features are
captured by an effective time-independent Bloch equation which provides an
analytical solution.  Moreover, we determine the decay of these arcs as a
function of dissipation strength and temperature.
\end{abstract}

\date{\today}


\maketitle

\section{Introduction}

The spectrum of a bistable quantum system as a function of the detuning
typically forms avoided crossings.  In particular in the regime
between adiabatic following and non-adiabatic transitions, sweeping the
detuning can induce a complex tunneling dynamics.  Its archetype is
a two-level system with a sweep linear in time for which the probability
for non-adiabatic transitions is given by the famous Landau-Zener formula
\cite{Landau1932a, Zener1932a, Stueckelberg1932a, Majorana1932a}.
It predicts the splitting of the wavefunction into a superposition of the
adiabatic qubit states, which means that the avoided crossing acts like a beam
splitter.
Replacing the linear switching by an ac field, results in a series of
avoided crossings so that the wavefunction splits and recombines
repeatedly---the quantum mechanical analogue of a Mach-Zehnder
interferometer \cite{Shevchenko2010a}.  The resulting LZSM
interference has been demonstrated in various experiments with solid-state
qubits \cite{Oliver2005a, Sillanpaa2006a, Wilson2007b, Berns2008a,
Stehlik2012a, Dupont-Ferrier2013, Li2013a}.

Going beyond the mere demonstration of interference, LZSM interferometry can be
employed as a tool to determine the dephasing time of a charge qubit.  The
analysis of the interference pattern may be performed in ``real space'', i.e.\
as a function of detuning and amplitude \cite{Dupont-Ferrier2013}, or in Fourier
space \cite{Forster2014a}.  The latter type of analysis is based on the
observation that the Fourier transform of LZSM patterns exhibit arc
structures with a characteristic decay \cite{Rudner2008a}.  By comparing measured
and computed patterns for a qubit, one can determine the inhomogeneous broadening
as well as the faster decoherence induced by substrate phonons
\cite{Forster2014a}.  Since this procedure takes considerable numerical effort,
any analytic knowledge may be helpful.

In this work, we reveal how the qubit-bath coupling operator and the shape
of the driving influence the LZSM interference pattern.  In
Sec.~\ref{sec:model} we describe the qubit as time-dependent spin-boson
model \cite{Grifoni1998a} and introduce the Floquet-Bloch-Redfield
formalism that provides our numerical solutions.
Section~\ref{sec:realspace} is devoted to the LZSM pattern in real space
which is governed by the coupling operator to the bath.  In
Sec.~\ref{sec:fourier} we demonstrate that its Fourier transform, by
contrast, mainly depends the shape of the driving.  Finally in
Sec.~\ref{sec:decay}, we determine the decay of the arcs as a function of
the bath parameters.

\section{Model and master equation}
\label{sec:model}

\subsection{Qubit in a time-dependent field}

We consider a qubit under the influence of a periodic driving
described by the Hamiltonian
\begin{equation}
\label{eq:Hamil1}
H(t)  = 
\begin{pmatrix}
\epsilon_0 & \Delta/2 \\
\Delta/2 & Af(t)
\end{pmatrix} ,
\end{equation}
where $\epsilon_0$ is a static detuning which is modulated by an ac driving
with amplitude $A$ and shape $f(t) = f(t+T)$.  The gauge chosen
in Eq.~\eqref{eq:Hamil1} is convenient for qualitative discussions, while
the equivalent symmetrized Hamiltonian $\tilde H(t) =
\frac{1}{2}\{\epsilon_0-Af(t)\} \sigma_z + \frac{\Delta}{2} \sigma_x$ is
preferable for the numerical treatment.

While the most prominent example is the monochromatic $f(t)=\cos(\Omega
t)$, our aim is to investigate LZSM interference for general periodic
driving.  In our numerical examples, we consider besides the purely
harmonic driving with $f_0(t)=\cos(\Omega t)$ also the shapes
\begin{subequations}
\label{eq:f}
\begin{align}
\label{eq:f1}
f_1(t) ={}& \cos(\Omega t) + 0.1\cos(3 \Omega t), \\
\label{eq:f2}
f_2(t) ={}& \cos(\Omega t) + \cos(2 \Omega t), \\
\label{eq:f3}
f_3(t) ={}& \sin(\Omega t) + \sin(2 \Omega t),
\end{align}
\end{subequations}
where $f_1$ and $f_2$ are symmetric functions, i.e., they obey
$f(t_0+t)=f(t_0-t)$ for $t_0=0$ and for $t_0=T/2$.  By contrast,
$f_3(t_0+t)=-f_3(t_0-t)$ is anti-symmetric.  While $f_1$ modifies
the pure cosine driving only slightly, the other two shapes are
qualitatively different because they possess several maxima and minima per
driving period. As we discuss below in Sec.~\ref{sec:fourier} this has
consequences for the structures observed in Fourier space,
see Fig.~\ref{fig:1}[(g)--(i)].

\subsection{System-bath model}
\label{sec:SBmodel}

The influence of the environment on the system is modeled by a bath of
harmonic oscillators given by the Hamiltonian 
$ H_\text{env} =H_{b}+H_\text{int} $
with
$ H_{b}=\sum_{\nu} \hbar\omega_{\nu} a_{\nu}^{\dagger}a_{\nu}, $
%
and
\begin{equation}
H_\text{int} = \frac{1}{2}X
\sum_{\nu}\hbar \lambda_{\nu} (a_{\nu}^{\dagger}+a_{\nu}),
\label{eq:H_int}
\end{equation}
where $\omega_{\nu}$ are the frequencies of the oscillators, while
$\hbar\lambda_{\nu}$ are the system-oscillator coupling energies. For the
qubit operator $X$ that couples to the bath, we consider
$\sigma_x$ and $\sigma_z$ as well as a linear combination of the two.
According to their orientation on the Bloch sphere with respect to the
driving, we refer to the coupling as transverse ($\sigma_x$) and longitudinal ($\sigma_z$), respectively.
Moreover we assume that system and environment are initially
uncorrelated, i.e., we choose an initial condition of the Feynman-Vernon type,
$\rho_\text{tot}(t_0)=\rho(t_0)\otimes R_\text{env,eq}$ for the total system density
operator $\rho_\text{tot}$, with $\rho(t_0)$ being the initial reduced density
operator of the qubit, while $R_\text{env,eq} \propto \exp(-\beta H_b)$ is the
Gibbs state of the bath with inverse temperature $\beta=1/k_B T$.

Starting from the Liouville-von Neumann equation $i \hbar \dot
\rho_\text{tot} = [H(t)+H_\text{env},\rho_\text{tot}]$ for the total
density matrix and applying standard techniques, one can derive the
Markovian weak-coupling master equation \cite{Kohler1997a}
\begin{align}
 \label{eq:master}
\frac{d}{dt} \rho &=\mathcal{L}(t) \rho \\
                  &=-i [H(t),\rho]  - \frac{1}{4} \int_0^{\infty} \hspace{-0.5cm}&d\tau\Bigl(& \mathcal{S}(\tau) [X,[\tilde X(t-\tau,t),\rho]] \notag\\  
                  & &+&\mathcal{A}(\tau)  [X,\{\tilde X(t-\tau,t),\rho\}] \Bigr) \notag,
\end{align}
where $\{A,B\}=AB +BA$ denotes the anti-commutator and $\tilde X(t',t)$ is
a shorthand notation for $U^{\dagger}(t,t') X U(t,t')$, with $U$ being the
propagator for the coherent qubit dynamics.  The influence of the
environment is subsumed in the symmetric and the antisymmetric bath
correlation function,
\begin{align}
 \label{eq:sym}
  \mathcal{S}(\tau) &= \frac{1}{2} \langle \{ B(\tau), B(0) \}\rangle_{\text{eq}} \notag\\
                 &=\frac{1}{\pi} \int_0^{\infty} d\omega J(\omega) \text{coth}(\hbar\omega\beta/2)\cos(\omega\tau),\\[10pt]
                 \label{eq:antisym}
  \mathcal{A}(\tau) &= \frac{1}{2} \langle [B(\tau), B(0) ]\rangle_{\text{eq}} \notag\\
                 &=\frac{1}{\pi} \int_0^{\infty} d\omega J(\omega) \sin(\omega\tau),
\end{align}
respectively, with the collective bath coordinate $B(t)=\sum_{\nu}\lambda_{\nu}
\{a_{\nu}^{\dagger} \exp(i\omega_{\nu} t)+a_{\nu}\exp(-i\omega_{\nu} t)\}$.
The angular brackets $\langle \dots \rangle_{\text{eq}}$ denote the average with
respect to the thermal equilibrium of the environment.  In a continuum
limit we consider the Ohmic spectral density $J(\omega) =  \pi \sum_{\nu}
\lambda^2_{\nu} \delta(\omega - \omega_{\nu}) \equiv 2\pi \alpha \omega
e^{-\omega_c/\omega} $ with the high-frequency cutoff $\omega_c$ eventually
taken to infinity.

\subsection{Bloch-Redfield theory in Floquet basis}

Since the system Hamiltonian is periodic in time we can apply the Floquet
theorem which states that the corresponding Schr{\"o}dinger equation
possesses a fundamental set of solutions of the form
$\ket{\Psi_{\alpha}(t)} = e^{-i\varepsilon_{\alpha}t/\hbar}
\ket{\Phi_\alpha(t)}$, with the quasi-energies $\varepsilon_{\alpha}$ and
the Floquet states $\ket{\Phi_\alpha(t)}=\ket{\Phi_\alpha(t+T)}$
\cite{Grifoni1998a}.  They can be calculated from the eigenvalue equation
$\{H(t)-i\hbar \partial_t\} \ket{\Phi_{\alpha}(t)} =
\varepsilon_{\alpha}\ket{\Phi_{\alpha}(t)}$. Expressing the master equation
Eq.~\eqref{eq:master} in the Floquet basis $\{ \ket{\Phi_{\alpha}(t)}\}$
\cite{Kohler1997a,Grifoni1998a} yields
\begin{equation}
 \label{eq:masterFl}
\frac{d}{dt} \rho_{\alpha \beta}(t) = 
 \sum_{\alpha'\beta',k}e^{-i k \Omega t}\mathcal{L}^{(k)}_{\alpha\beta, \alpha'\beta'} \rho_{\alpha'\beta'}(t).
\end{equation}
with the density matrix element $\rho_{\alpha\beta}= \bra{\Phi_{\alpha}(t)}
\rho \ket{\Phi_{\beta}(t)}$ and
\begin{align}
 \label{eq:Liouville}
\mathcal{L}^{(k)}_{\alpha \beta, \alpha'\beta'}
= &-i(\varepsilon_{\alpha'} - \varepsilon_{\beta'})
  \delta_{\alpha,\alpha'}\delta_{\beta,\beta'}\delta_{0,k} \\ 
  &+ \sum_{k'} (N_{\alpha\alpha',k'} +N_{\beta\beta',k'-k} )
   X_{\alpha\alpha',k'} X_{\beta'\beta,k-k'} \notag \\ 
  &+ \delta_{\beta,\beta'}\sum_{k',\beta''} N_{\beta'' \alpha', k-k'}
   X_{\alpha\beta'',k'} X_{\beta''\alpha',k-k'}\notag\\
  &+ \delta_{\alpha,\alpha'}\sum_{k',\alpha'} N_{\alpha'' \beta', k'-k}
   X_{\beta'\alpha'',k-k'} X_{\alpha''\beta,k'} \notag \,,
\end{align}
the $k$th Fourier coefficients of the Liouville operator.  We introduced the
transition matrix elements
\begin{equation}
	X_{\alpha\beta,k}= \frac{1}{T}\int_0^{T} dt \, e^{i k
\Omega t} \langle\Phi_{\alpha}(t)| X |\Phi_{\beta}(t)\rangle
\end{equation}
and
$N_{\alpha\beta,
k}=N(\varepsilon_{\alpha}-\varepsilon_{\beta}+k\hbar\Omega)$ with
$N(\omega)=\alpha \omega  n_{\text{th}}(\omega) $ and the bosonic thermal
occupation number $n_{\text{th}}(\omega)=(e^{\beta\hbar\omega}-1)^{-1}$. 

In the long-time limit, the system relaxes to a steady state which obeys
the time-periodicity of the driving, $\rho_\infty(t)=\rho_\infty(t+T)$.
Hence we can use the Fourier decomposition  $\rho_{\infty}(t) = \sum_k
e^{-i k \Omega t} \rho^{(k)}$ of the density operator to obtain
\begin{equation}
 \label{eq:linearsys}
-i\hbar k\Omega \rho_{\alpha\beta}^{(k)}
= \sum_{\alpha',\beta',k'} 
  \mathcal{L}_{\alpha\beta,\alpha'\beta'}^{(k-k')}\rho_{\alpha'\beta'}^{(k')} .
\end{equation}
This master equation avoids the common moderate \cite{Kohler1997a} or full
\cite{Blumel1991a} rotating-wave approximation with respect to the driving
frequency $\Omega$ and, hence, is rather reliable \cite{Lehmann2002b}.  In
our case, numerical convergence is already obtained with $|k|\leq 5$, i.e.,
for truncation at the fifth sideband, even when the Floquet states may
contain many more relevant sidebands.  Thus the numerical effort for
solving Eq.~\eqref{eq:linearsys} stays at a tolerable level.

\begin{figure*}
\centerline{\includegraphics[]{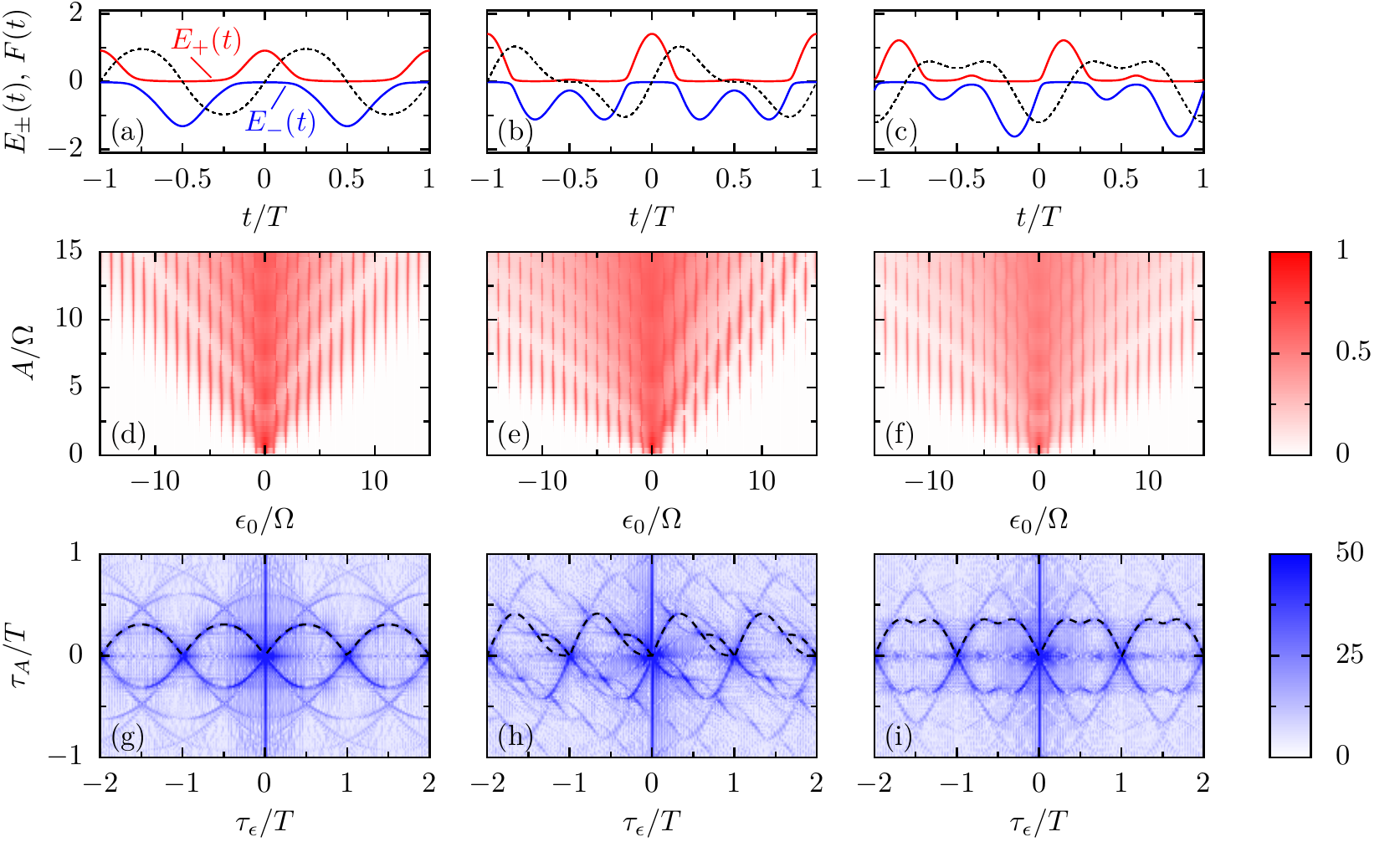}}
\caption{(Color online)
[(a)--(c)] Adiabatic energies $E_\pm(t)$ (red and blue solid lines) of the Hamiltonian
\eqref{eq:Hamil1} in units of $\hbar\Omega$ for vanishing static detuning,
$\epsilon_0=0$, and the driving shapes $f_1(t)$--$f_3(t)$ in Eq.~\eqref{eq:f}.  The dashed
black line marks the integral of the driving, $F(t)$, in units of
$1/\Omega$.  [(d)--(f)] Resulting non-equilibrium populations in
$\epsilon_0$-$A$ space.  [(g)--(i)] 2D Fourier transform $W(\tau_\epsilon,\tau_A)$ of the
interference patterns, defined in Eq.~\eqref{eq:Foutrafo}.  The dashed lines in the upper half plane mark the
analytic expressions for the arc structure derived in
Sec.~\ref{sec:fourier}.  The patterns are computed with the stationary
solution of the Bloch-Redfield master equation for the tunnel matrix
element $\Delta=0.5\Omega$, dissipation strength $\alpha=10^{-3}$,
temperature $k_BT= 1/\beta=0.1\hbar\Omega$, and transverse qubit-bath
coupling, i.e., $X=\sigma_x$ in Eq.~\eqref{eq:H_int}.
\label{fig:1}}
\end{figure*}

\subsection{Excitation probability}

For the visualization of the LZSM interference pattern, one may in the
absence of dissipation consider time-averaged transition probabilities from
a particular initial state \cite{Shevchenko2010a}. In the presence of a
heat bath, however, the system state is in the long-time limit typically
independent of the initial state. Therefore, we consider
time averages of observables such as populations, e.g., of the diabatic state 
$|{\uparrow}\rangle$, or the excited state of the undriven qubit, $|e\rangle$. Since in the vast part of the parameter space considered, the
qubit is strongly biased, i.e., $\Delta\ll|\epsilon_0|$, the choice is of
minor practical relevance.  We here consider the latter namely the
time-averaged probability for the qubit being in the excited state as a
function of the static detuning $\epsilon_0$ and the driving amplitude $A$,
\begin{equation}
P_\text{ex}(\epsilon_0,A)
= \frac{1}{T} \int_0^T dt\, \langle e|\rho_\infty(t) |e\rangle ,
\end{equation}
where $\rho_{\infty}(t)$ is the periodic long-time solution of the master
equation. Thus, $P_\text{ex}$ directly relates to the Fourier coefficients
in Floquet basis, $\rho_{\alpha\beta}^{(k)}$.

\section{Interference pattern in energy space}
\label{sec:realspace}

In order to give a first impression of our results, we depict in
Figs.~\ref{fig:1}(d)--(f) the LZSM interference patterns for the driving
shapes in Eq.~\eqref{eq:f} and transverse qubit-bath coupling.  All three
patterns exhibit resonance peaks whenever the detuning $\epsilon_0$ matches
with a multiple of the driving frequency.  As a further condition for a
significant non-equilibrium population, the amplitude must be so large that
it reaches the avoided level crossing, which is the case for $\min[f(t)] <
\epsilon_0/A < \max[ f(t)]$. The peaks depend strongly on the amplitude and
may even vanish.  This represents a generalization of coherent destruction
of tunneling found for sinusoidal driving \cite{Grossmann1992a}, a
phenomenon responsible for the characteristic vertical structure of LZSM
patterns \cite{Shevchenko2010a} which, in turn, can be explained within a
Landau-Zener scenario \cite{Kayanuma1994a}.  Comparing panels (d)--(f), we
can conclude that the patterns look qualitatively the same, despite the
rather different driving shapes which are visible in the adiabatic energies
of the qubit Hamiltonian \eqref{eq:Hamil1} depicted in panels (a)--(c).
The main differences stem from the fact that the harmonics with frequency
$2\Omega$ may change the maximum and the minimum value of $f(t)$ and, thus,
affect the above condition for significant excitations.  For the driving shape
$f_2$, this condition depends on the sign of $\epsilon_0$, which explains
the asymmetry of the pattern in panel (e), which was also observed in
Ref.~\cite{Satanin2014a}.

\subsection{Influence of the qubit-bath coupling}

\begin{figure}
\includegraphics{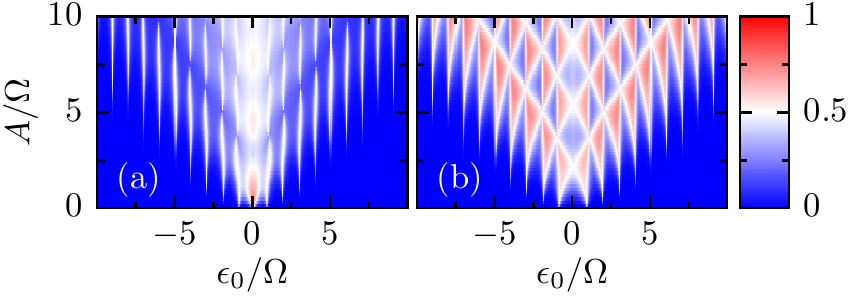}
\caption{(Color online) Non-equilibrium population $P_\text{ex}$ as a function of the
detuning $\epsilon_0$ and the driving amplitude $A$ for $f(t)=\cos(\Omega
t)$.  The qubit-bath coupling $H_\text{int}$ is determined by $X=\sigma_x$ (a) and
$X=\sigma_z$ (b), while $\Delta=0.5\Omega$, $\alpha=10^{-3}$ and
$1/\beta=0.1\hbar\Omega$.}
\label{fig:P(e,A)}
\end{figure}
In Fig.~\ref{fig:P(e,A)}, we compare patterns for transverse and
longitudinal qubit bath coupling, i.e., the coupling via $\sigma_x$ and
$\sigma_z$, respectively.  Since we already noticed that the pattern in
energy space is not very sensitive to the shape of the driving, we here
restrict ourselves to the purely harmonic $f_0(t)=\cos(\Omega t)$.  Let us
first consider the transverse coupling.  The resulting pattern
[Fig.~\ref{fig:P(e,A)}(a)] is characterized by resonance islands which as a
function of the detuning $\epsilon_0$ are Lorentzians.  As a function of
the amplitude $A$, their shape follows approximately the squares of Bessel
functions.  This behavior was predicted for the current through ac-gated
double quantum dots \cite{Strass2005b, Forster2014a} and for the
non-equilibrium population of a driven two-level system
\cite{Shevchenko2010a}.  Moreover, it has been observed with good
resolution in various experiments \cite{Oliver2005a, Wilson2007b,
Izmalkov2008a, Rudner2008a, Stehlik2012a, Forster2014a}.

If the bath couples longitudinally with respect to the driving, i.e., when
both the ac field and the environment enter via $\sigma_z$, the pattern
changes qualitatively.  As can be appreciated in Fig.~\ref{fig:P(e,A)}(b),
the Lorentizan peaks turn into a triangular structure.
This kind of bath coupling should be relevant for a charge qubit in a
Cooper pair box driven by an ac gate voltage while being sensitive to
environmental charge fluctuations.  The LZSM pattern for such a case has
been measured in Ref.~\cite{Sillanpaa2006a} and indeed exhibits some
similarity with Fig.~\ref{fig:P(e,A)}(b).  However, the resolution of the
experimental data is not sufficient for an unambiguous comparison.

\begin{figure}
\includegraphics[]{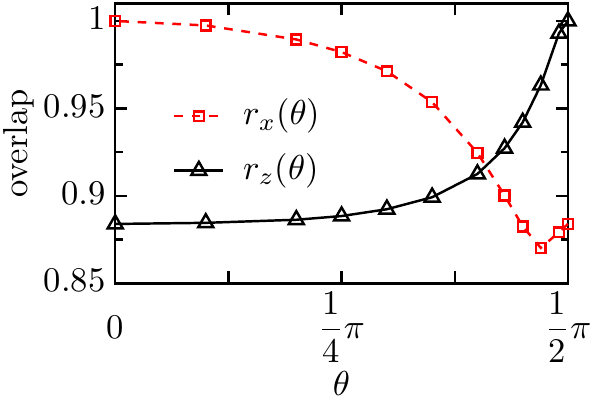}
\caption{Overlap of the interference pattern for the mixed coupling
\eqref{mixed.coupling} with the patterns for the coupling operators
$\sigma_x$ (squares) and $\sigma_z$ (triangles) as a function of the mixing
angle.  All other parameters are as in Fig.~\ref{fig:P(e,A)}.
}
\label{fig:overlap}
\end{figure}

As a generalization of these two system-bath couplings, we also
considered the coupling via the operator
\begin{equation}
\label{mixed.coupling}
X = \sigma_x\cos\theta+\sigma_z\sin\theta .
\end{equation}
The mixing angle $\theta$ varies from $0$  to $\pi/2$, where the limits
$\theta=0$ and $\theta=\pi/2$ correspond to the transverse and
the longitudinal case.  This model captures, e.g., a superconducting charge
qubit that interacts capacititvely as well as inductively with the
environmental circuitory.  Then it is intriguing which dissipative
influence dominates the LZSM interference.
For this purpose, we define via the inner product of the patterns in
parameter space the normalized overlaps $r_{x,z}(\theta)$ of the
$\theta$-dependent pattern with those for the bath couplings via
$\sigma_x$ and $\sigma_z$.  Obviously, their limits are $r_x(0) = 1 =
r_z(\pi/2)$.

The result shown in Fig.~\ref{fig:overlap} reveals that upon increasing
$\theta$ from $\theta=0$, i.e., augmenting the influence of $\sigma_z$, the
pattern remains close to the one of Fig.~\ref{fig:P(e,A)}(a).  By contrast,
the pattern for $\sigma_z$ coupling is more sensitive to a small admixture
of $\sigma_x$.  Thus, unless the bath coupling via $\sigma_z$ is much
larger, we find the ``usual'' interference pattern of
Fig.~\ref{fig:P(e,A)}(a).  This is consistent with the fact that in most
experiments, one indeed finds such a LZSM pattern with Lorentzians
\cite{Oliver2005a, Wilson2007b, Izmalkov2008a, Rudner2008a}.  Notice
however that this reasoning does not necessarily apply to LZSM patterns for
the average current through open double quantum dots \cite{Stehlik2012a,
Forster2014a}, because there the dominating incoherent dynamics is the
electron tunneling between the quantum dots and the leads.  Moreover, the
Hilbert space for a transport setup is larger since it must comprise states
with different electron number.

Let us emphasize that the observed significant dependence of the long-time
solution on the coupling is found even in the limit of very weak qubit-bath
coupling and, hence, it is beyond a mere higher-order effect in the
dissipation strength $\alpha$.
This is in clear contrast to the stationary solution of the
Bloch-Redfield equation for a time-independent problem, which generally is
the grand canonical state, while possible deviations are of the order
$\alpha$ \cite{Thingna2012a}.  Nevertheless, we will be able to derive an
effective time-independent Bloch equation for the driven qubit which
captures the influence of the bath coupling operator quantitatively.

\subsection{Analysis of the resonance peaks}
\label{sec:PeaksAnalytical}

\begin{figure}
\includegraphics{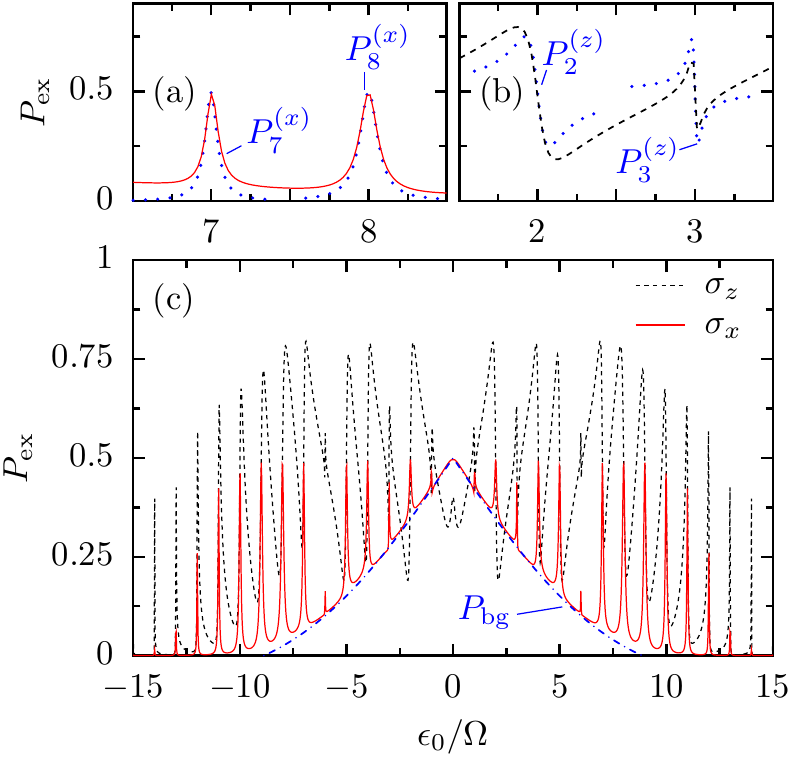}
\caption{(Color online)
Non-equilibrium population $P_\text{ex}$ shown in Fig.~\ref{fig:P(e,A)} as
a function of the detuning $\epsilon_0$ for the driving amplitude $A=10\Omega$.
(a) Comparison between numerical result with $\sigma_x$ coupling obtained
with the Bloch-Redfield master equation (solid red) and the analytical
solution \eqref{Px} for the resonances with $n=7,8$ (dotted blue).
(b) Comparison between numerical result with $\sigma_z$ coupling (dashed
black) and the analytical solution \eqref{Pz} for n=2,3 (dotted blue).
(c) Numerical results for $\sigma_x$ and $\sigma_z$ coupling plotted
together with the analytical result for the off-resonant background
predicted by Eq.~\eqref{Pbg}.}
\label{fig:P(e)}
\end{figure}
While the two-dimensional interference patterns in
Figs.~\ref{fig:1}(d)--(f) and \ref{fig:P(e,A)} provide a comprehensive
picture, the details of the resonance peaks are better visible in the
horizontal slices shown in Fig.~\ref{fig:P(e)}.  They reveal that the peaks
for transverse coupling indeed are Lorentzians.  For longitudinal coupling,
the peaks are anti-symmetric.  Moreover, we witness a triangular shaped
background with a roughly linear decays in $|\epsilon_0|$ while being
practically independent of the tunneling $\Delta$. Our aim is to explain
these features within the Bloch equations for the qubit derived in
Appendix~\ref{app:Bloch}.  We restrict the discussion to the limit of very
low temperatures for which the interference pattern is most pronounced.

\subsubsection{Off-resonant background}
\label{sec:back}

We start our considerations by noticing that at low temperatures, the
dissipative dynamics is mainly a decay towards the qubit ground state.
Since for small tunneling $\Delta$ and large amplitude $A$, the (adiabatic)
qubit levels form avoided crossings, the states $|{\downarrow}\rangle$ and
$|{\uparrow}\rangle$ take turns in having lower energy, cf.\ the upper row of
Fig.~\ref{fig:1}.  Within an adiabatic description, we employ the Bloch
equation \eqref{eq:Bloch} and replace the $\epsilon$-dependent rates by
their instantaneous value to obtain for the $z$-component of the Bloch
vector $\vec s = \tr(\vec\sigma\rho)$ the equation of motion
\begin{equation}
\dot s_z = -\Gamma[\epsilon(t)]s_z - \pi\alpha\epsilon(t) ,
\end{equation}
where $\epsilon(t) = \epsilon_0+Af(t)$.  If the decay is sufficiently slow,
we can replace the time-dependent coefficients by their time averages
$\bar\Gamma \equiv \overline{\Gamma[\epsilon(t)]} \approx
\alpha(2A+\epsilon_0^2/A)$ and $\overline{\epsilon(t)} = \epsilon_0$.  Then
the steady-state solution $s_z = \pi\alpha\epsilon_0/\bar\Gamma$
corresponds to the non-equilibrium population
\begin{equation}
\label{Pbg}
P_\text{bg} = \frac{1}{2} - \frac{\pi\epsilon_0 A}{4A^2+2\epsilon_0^2} .
\end{equation}
The dashed-dotted line in Fig.~\ref{fig:P(e)} (c) shows that this estimate
indeed describes the triangular shaped background rather well which, in
turn, confirms the underlying adiabatic picture.

\subsubsection{Lorentzian peaks for the transverse coupling via $\sigma_x$}
\label{sec:Pnx}

An analytic expression for the resonance peaks can be found within an
approximation scheme for close-to-resonant excitation \cite{Strass2005b,
Shevchenko2010a}.  For a bath coupling via $\sigma_x$, the calculation is
practically the one given in the appendix of Ref.~\cite{Shevchenko2010a}.
We sketch it briefly so that we can later highlight the differences to the
case of a bath coupling via $\sigma_z$.

Embarking with the master equation~\eqref{eq:master} we consider the limit $\epsilon_0
\gg \Delta$ and assume that the driving frequency is close to resonance,
i.e., $n\Omega= (\epsilon_0^2+\Delta^2)^{1/2}\approx \epsilon_0 $.
In this regime the tunneling contribution, proportional to $\Delta$
represents a perturbation to the free dynamics governed by
$\frac{1}{2}\{\epsilon_0+f(t)\}\sigma_{z}$.  In order to capture the
coherent dynamics in large part, we apply the unitary
transformation $U(t) = \exp\{-i\phi(t)\sigma_z/2\}$ with the time-dependent
phase $\phi(t) = n\Omega t +A F(t)$, where
\begin{equation}
F(t) = \int_0^t dt'\, f(t')
\end{equation}
obeys the $T$-periodicity of the driving since $f(t)$ by definition
vanishes on average.  Then we obtain the interaction
picture Hamiltonian $U^{\dagger}(t)H(t)U(t)-i\hbar U^{\dagger}(t)
\dot{U}(t)$.  Averaging over the driving period $T$ results in the
effective Hamiltonian
\begin{equation}
 \label{eq:effective2}
  H_\text{eff} = -\frac{\delta_n}{2} \sigma_z + \frac{\Delta_n}{2} \sigma_x,
\end{equation}
with the detuning $\delta_n = n\hbar\Omega - \epsilon_0$ and the effective
tunnel matrix element 
\begin{equation}
 \label{eq:Delta2}
\Delta_n(A) = \frac{\Delta}{T} \int_0^T dt\, e^{in\Omega t-iAF(t)} .
\end{equation}
The latter obviously is the $n$th Fourier coefficient of
$\Delta\exp\{-iAF(t)\}$, a property that will prove useful.  This
generalizes the result for purely harmonic driving, $\Delta_n = \Delta
J_n(A/\Omega)$ with the $n$th order Bessel function of the first kind, to
arbitrary but periodic shapes $f(t)$.
The corresponding equation of motion for the Bloch vector reads
$\dot{\vec{s}} = \vec B_\text{eff} \times\vec s$, where $\vec B_\text{eff}
= (\Delta_n,0,-\delta_n)^T$.

For the dissipative dynamics, we distinguish two limiting cases.  First, during
the stage at which the qubit passes through the crossing, the tunneling term
$\Delta\sigma_x/2$ dominates in the Hamiltonian \eqref{eq:Hamil1}, while the
qubit-bath coupling essentially commutes with the Hamiltonian.  Thus, it induces
pure dephasing but no decay.  Since for an Ohmic bath, the dephasing rate
\eqref{app:Gammaphi} is proportional to the temperature, it can be
neglected in the limit under consideration.

For most of the time, however, the qubit Hamiltonian is dominated by the
term proportional to $\sigma_z$ so that the bath causes transitions between
the eigenstates $H_\text{eff}$.  We describe them by the Bloch equation
\eqref{eq:Bloch} which together with the effective coherent dynamics reads
\begin{equation}
\label{Bloch}
\dot{\vec{s}}
=\begin{pmatrix}-\Gamma/2 & -\delta_n & 0 \\
                 \delta_n &- \Gamma/2& \Delta_n \\
                0 & -\Delta_n & -\Gamma
  \end{pmatrix} \vec s
 -\begin{pmatrix} 0 \\ 0 \\ \Gamma \end{pmatrix} .
\end{equation}
Notice that since we are only interested in the stationary state,
we can ignore possible driving-induced
renormalizations of the decay rates \cite{Fonseca2004a} and treat $\Gamma$
as phenomenological parameter.  However, we like to stress that our
numerical treatment captures this renormalization.  The steady state $\vec
s(\infty)$ is easily obtained by matrix inversion and provides the
non-equilibrium population
\begin{equation}
\label{Px}
P_n^{(x)} = \frac{1}{2}
\frac{\Delta_n^2/2}{(\epsilon_0-n\Omega)^2+\Delta_n^2/2+\Gamma^2/4} .
\end{equation}
While this expression holds close to the $n$th resonance, it vanishes
far-off.  Therefore, the global picture is simply given by the sum of the
contributions of all resonances and reads $P_\text{ex}^{(x)} =
\sum_n P_n^{(x)}$.  Such expressions have been found
not only for non-equilibrium populations of driven qubits
\cite{Rudner2008a, Shevchenko2010a} but also for the dc current through
double quantum dots \cite{Strass2005b, Forster2014a}.

In Fig.~\ref{fig:P(e)}(a), we compare the numerically computed interference
pattern for the $\sigma_x$ coupling with the analytical solution \eqref{Px}
at various resonances. While close to the resonances, i.e.\ for $\delta_n
\ll\Delta$, the agreement is almost perfect, we observe small deviations
between the resonances which mainly stem from the off-resonant background
discussed above.

\subsubsection{Anti-symmetric resonances for the longitudinal coupling via $\sigma_z$}

For longitudinal coupling, the situation is complementary to the transverse
case.  Outside the crossing, the bath couples to a good quantum number
of the qubit and, thus, creates pure dephasing negligible at low
temperatures.  Thus, dissipative transitions are only induced close to the
crossing.  Therefore we obtain the corresponding Bloch equations by cyclic
permutation of the dissipative terms in Eq.~\eqref{Bloch} which yields
\begin{equation}
\label{BlochZ}
\dot{\vec{s}}
=\begin{pmatrix}-\Gamma & -\delta_n & 0 \\
                 \delta_n &- \Gamma/2& \Delta_n \\
                0 & -\Delta_n & -\Gamma/2
  \end{pmatrix} \vec s
 -\begin{pmatrix} \Gamma \\ 0 \\ 0 \end{pmatrix} .
\end{equation}
Its stationary solution provides the non-equilibrium population
\begin{equation}
\label{Pz}
P_n^{(z)}
= \frac{1}{2} + \frac{(\epsilon_0-n\Omega)\Delta_n}
  {(\epsilon_0-n\Omega)^2+2\Delta_n^2+\Gamma^2/2}.
\end{equation}
Since now the qubit decay occurs only during the short stages when
the levels cross, the phenomenological rate $\Gamma$ is expected to be
considereably smaller than for $\sigma_x$ coupling.

In Fig.~\ref{fig:P(e)}(b), we compare the numerically computed interference
pattern obtained with $\sigma_z$ coupling with the analytical solution
\eqref{Pz} for $n=2,3$.  Again, close to a resonance the analytics and the
numerical solution agree rather well.  Far from resonance, however,
expression \eqref{Pz} decays only slowly and, the global picture is beyond
the simple summation of all $P_n^{(z)}$.
Exactly on resonance, i.e., for $\epsilon_0 = n\Omega$,
the second term of Eq.~\eqref{Pz} vanishes and, hence, the excitation
probability becomes $P_\text{ex}^{(z)} = 1/2$ for all $n$.
This fact together with the asymmetry of the structure implies that close
to each resonance, we find a region with $P_\text{ex}^{(z)}>1/2$.
Such population inversion has been found also for driven qubits with other
structureless bath spectral densities \cite{Stace2005}.
 
\section{Interference pattern in Fourier space}
\label{sec:fourier}

While we found that the interference patterns in real space depend only weakly
in the shape of the diving, the opposite is true for their 2D
Fourier transform shown in Figs.~\ref{fig:1}(g)--(i).
For the symmetric driving functions $f_1$ and $f_2$, we find a pronounced
arc structure at $\tau_A = 2F(\tau_\epsilon/2)$ and $\tau_A =
2F(\tau_\epsilon/2+T/2)$, cf.\ the dashed black lines in
panels (a) and (b). They can be explained within the stationary-phase
treatment of the LZSM interference scenario \cite{Rudner2008a}.  However,
there emerge several features that are beyond.  Most
significantly in panel (i), we find that for the anti-symmetric driving
with $f_3$, the structure is different from the corresponding
$F(\tau_\epsilon/2)$ depicted by the dashed line in panel (c).  Moreover,
the driving $f_2$ yields additional arcs close to the origin.  There also
emerge higher-order replica of the arcs which have been found both
experimentally \cite{Rudner2008a, Forster2014a} and theoretically
\cite{Forster2014a}.

For an analytical approach to the arc structure, we consider $
P_\text{ex}(\epsilon_0,A) = \sum_n P_n^{(x)}(\epsilon_0,A) $ derived
Sec.~\ref{sec:Pnx} and define its Fourier transform as
\begin{equation}
W(\tau_{\epsilon},\tau_{A}) =
\int \frac{d\epsilon_0}{2\pi} \, \frac{dA}{2\pi}\, e^{-i \epsilon_0
\tau_{\epsilon}}e^{-i A \tau_{A}} P_\text{ex}(\epsilon_0,A) .
\label{eq:Foutrafo}
\end{equation}
The
$\epsilon_0$-integral can be evaluated readily to yield
\begin{equation}
\label{eq:fou1}
W(\tau_\epsilon,\tau_A) = \frac{1}{4\pi} \int dA\,
  e^{-iA\tau_A} \sum_n \frac{\Delta_n^2}{\Gamma_n^*}
  e^{-in\Omega\tau_\epsilon} e^{-\Gamma^*_n|\tau_\epsilon|},
\end{equation}
with the resonance width $\Gamma_n^*=(\Delta_n^2/2+\Gamma^2/4)^{1/2}$.

\subsection{Overdamped limit}

The remaining $A$ integral in \eqref{eq:fou1} can be evaluated directly in
the over-damped limit $\Gamma \gg \Delta$ in which
$\Gamma^*_n \approx \Gamma/2$ and, thus,
\begin{equation}
\label{eq:overdamped}
W(\tau_\epsilon,\tau_A) = \frac{1}{2\pi\Gamma} \int dA e^{-iA\tau_A} \sum_n
\Delta_n^2  e^{-in\Omega\tau_\epsilon} .
\end{equation}
Focusing on the range of small $\tau_\epsilon$, we have neglected the last
exponential in Eq.~\eqref{eq:fou1}.
To proceed, we evaluate the sum
\begin{equation}
\label{sumDelta2}
\sum_n \Delta_n \cdot \Delta_n e^{-in\Omega \tau_\epsilon} ,
\end{equation}
where the two factors are easily identified as the $n$th Fourier
coefficients of $\exp\{-iAF(t)\}$ and $\exp\{-iAF(t+\tau_\epsilon)\}$,
respectively, cf.\ the definition of $\Delta_n$ in Eq.~\eqref{eq:Delta2}.
Thus, expression \eqref{sumDelta2} represents the inner product of these
exponentials.  According to Parseval's theorem, it can be written in the
time domain to read
\begin{equation}
\frac{1}{T} \int_0^T dt\, e^{iAF(t)} e^{-iAF(t+\tau_\epsilon)} .
\end{equation}
We symmetrize the integrand via the substitution $t\to t-\tau_\epsilon/2$
and perform the $A$-integration to obtain
\begin{align}
W(\tau_\epsilon,\tau_A)
={} & \frac{1}{T} \int_0^T dt\,
      \delta(\tau_A - G(t,\tau_\epsilon))
\label{WG}
\\
={} & \frac{1}{T} \sum_{t_i} \frac{1}{|g(t_i,\tau_\epsilon)|} ,
\label{Wg}
\end{align}
where
\begin{align}
G(t,\tau_\epsilon) ={}& F(t+\tau_\epsilon/2) - F(t-\tau_\epsilon/2) ,
\\
g(t,\tau_\epsilon) ={}& f(t+\tau_\epsilon/2) - f(t-\tau_\epsilon/2) .
\end{align}
The sum in Eq.~\eqref{Wg} has to be taken over all times $t_i$ that fulfill
$\tau_A=G(t_i,\tau_\epsilon)$.

\begin{figure}
\includegraphics[scale=1]{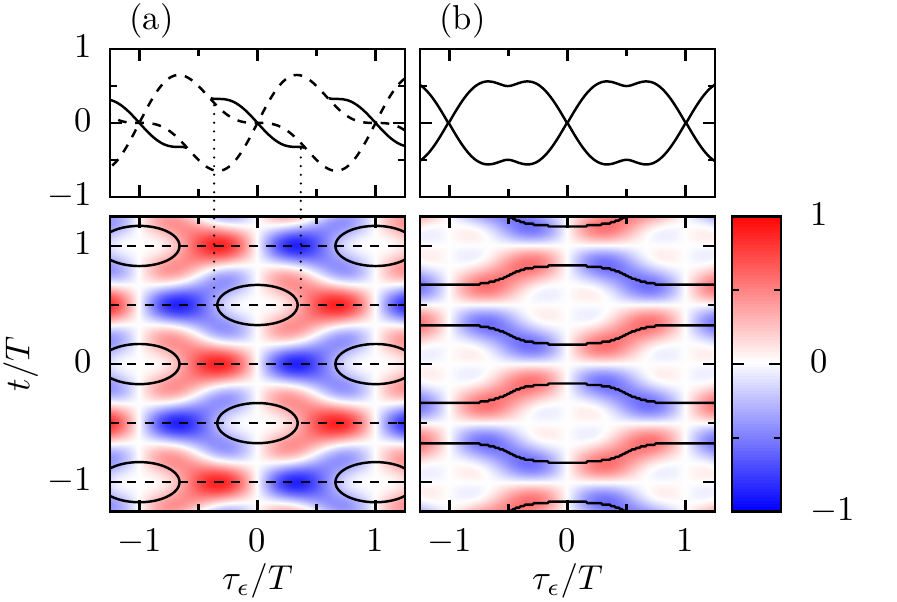}
\caption{(Color online)
Determination of the ``non-generic'' arcs for the driving shapes
$f_2$ (a) and $f_3$ (b).  The color code in the lower panels depicts
$G(t,\tau_\epsilon)$, while the horizontal dashed lines mark the generic
solutions of Eq.~\eqref{gzero} at multiples of $T/2$. The solid lines
represent numerical solutions of Eq.~\eqref{gzero}. Significant
contributions to $W(\tau_\epsilon,\tau_A)$ are determined by the solutions
of the transcendental equations \eqref{gzero} and \eqref{tauG}, i.e.\ the
cuts of $G(t,\tau_\epsilon)$ along the solid and dashed lines. Projecting
these solutions on the $\tau_\epsilon$ axis (hinted by vertical dotted lines)
results in the arc structures plotted in the upper panels and in
Figs.~\ref{fig:1}(h) and \ref{fig:1}(i).
}
\label{fig:2}
\end{figure}
Expressions \eqref{WG} and \eqref{Wg} allow us to extract the arc
structure by the following reasoning.  On the one hand, the argument of the
delta-function in Eq.~\eqref{WG} specifies the times $t_i$ that contribute
to the integral.  On the other hand, the most significant contributions to
$W$ stem from regions where the denominator in Eq.~\eqref{Wg} vanishes.
Thus, the structure is determined by the conditions
\begin{align}
\label{gzero}
0 ={}& g(t,\tau_\epsilon) , \\
\label{tauG}
\tau_A ={}& G(t,\tau_\epsilon),
\end{align}
which describe one-dimensional manifolds in the Fourier space
$(\tau_\epsilon,\tau_A)$.  They correspond to the arcs in
Figs.~\ref{fig:1}(g)--(i).  Practically, the arc structure is obtained in
the following way.  One determines from $g(t_i,\tau_\epsilon)=0$ all zeros
$t_i(\tau_\epsilon)$ and inserts them into Eq.~\eqref{tauG} which yields
relations of the type $\tau_A^{(i)}(\tau_\epsilon)$.

Obviously, $\tau_A = \tau_\epsilon = t = 0$
is a trivial solution for any driving shape $f(t)$.  Thus, the Fourier
transformed of all LZSM patterns exhibits a peak at the origin and, owing
to the periodcity of the driving, at multiples of $T$.

Two generic arcs can be found analytically if the driving obeys
time-reversal symmetry,
$f(t-t_s) = f(-t-t_s)$ (without loss of generality, we henceforth assume
$t_s=0$).  Then Eq.~\eqref{gzero} possesses the solutions $t_1=0$ and,
owing to the $T$-periodicity of $f$, $t_2=T/2$.  They provide the arcs
\begin{align}
\label{tauA1}
\tau_A^{(1)} ={}& 2F(\tau_\epsilon/2) , \\
\label{tauA2}
\tau_A^{(2)} ={}& 2F(\tau_\epsilon/2 +T/2) ,
\end{align}
which are in agreement with Ref.~\cite{Rudner2008a}.

If a symmetric driving $f$ has only one minimum and one maximum per period,
such as $f_1$ or $f(t)=\cos(\Omega t)$, $t_1$ and $t_2$ are the only
roots of Eq.~\eqref{gzero}.  Then the arc structure for symmetric
driving can be obtained fully analytically.  This fact is of practical use
if one employs LZSM interference to determine decoherence properties of a
qubit via the arc decay \cite{Forster2014a}.

In all other cases, i.e., when $f$ is not symmetric or if it possesses more
than two extrema per period, we have to solve Eq.~\eqref{gzero} numerically
to obtain also ``non-generic'' arcs.  For the symmetric driving $f_2$, this
leads to the ellipse-shaped solutions sketched in the lower panel of
Fig.~\ref{fig:2}(a).  Upon reducing the harmonic with frequency $2\Omega$,
they shrink and eventually vanish.  Together with the generic solution, we
obtain the structure shown in the upper panel of Fig.~\ref{fig:2}(a).  In
particular, there is a region in which the arc splits into two branches.
This prediction is quantitatively confirmed by the numerical solution of
the full problem shown in Fig.~\ref{fig:1}(h).

If $f$ is not time-reversal symmetric, we generally have to determine all $t_i$
numerically.  For the driving shape $f_3$, this procedure is visualized in
Fig.~\ref{fig:2}(b), where the solid lines in the lower panel depict the
zeros of $g(t,\tau_\epsilon)$ which define two independent manifolds
$t_i(\tau_\epsilon)$ and those related by the time shift $t\to t+T$.  The
corresponding arc structure shown in the upper panel agrees with the one
obtained numerically which is shown in Fig.~\ref{fig:1}(i).

\subsection{Weak dissipation}

In the limit of weak dissipation, $\Gamma\ll\Delta_n$, the resonance width
in Eq.~\eqref{eq:fou1} becomes $\Gamma_n^* = |\Delta_n|/\sqrt{2}$, so that
we have to evaluate the Fourier transform of $\sum_n|\Delta_n(A)|$.  This
represents a rather difficult task and, thus, we only discuss its
implications on a qualitative level.

A main effect of the cusps stemming from the absolute value is the
emergence of higher harmonics, cf.\ the Fourier transform of expressions
such as $|\cos(\Omega t)|$.  Accordingly, in the Fourier transform of our
interference patterns, we find arcs of higher order as can be appreciated
in Figs.~\ref{fig:1}(g)--(i).  To be specific, the arcs given by
Eqs.~\eqref{tauA1} and \eqref{tauA2} are generalized to
\begin{equation}
\tau_A = 2k F(\tau_\epsilon/2k + k'T/2k) ,
\end{equation}
where $k=1,2,3,\ldots$ and $k'=0,1,\ldots,2k-1$.  This prediction agrees
with our numerical findings shown in Figs.~\ref{fig:1}(g) and
\ref{fig:1}(h).  From a theoretical point of view, it is interesting to see
that arcs of higher order are found already within a two-level description,
i.e., within the most basic model for LZSM interference.  Thus, their
emergence does not require the consideration of further levels or
non-linearities.

\section{Decay of the arc structure}
\label{sec:decay}

\begin{figure}
\includegraphics[scale=1]{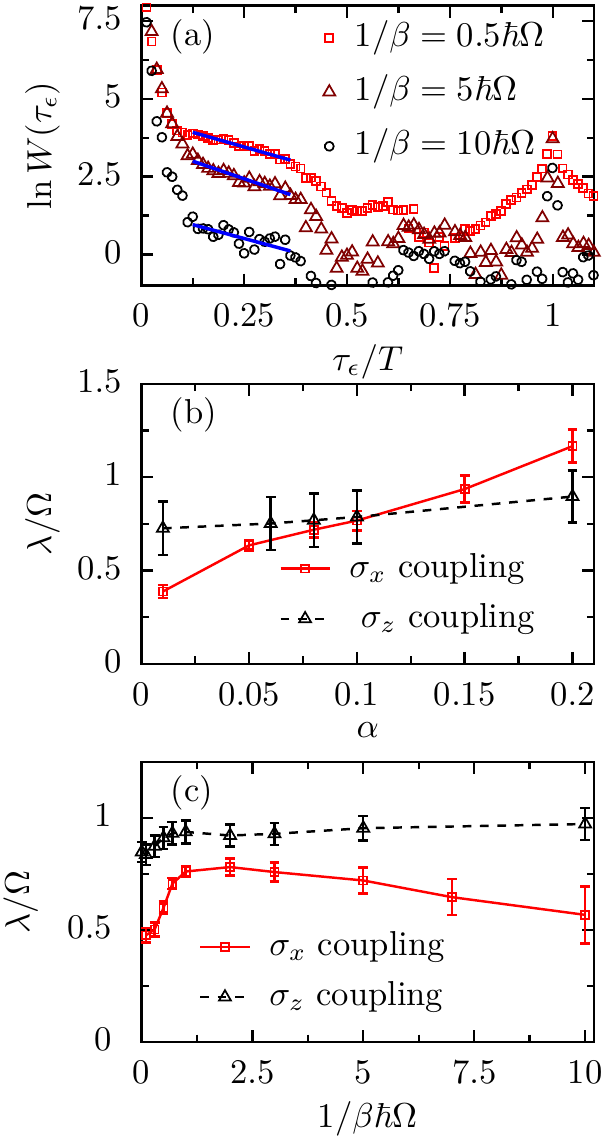}
\caption{(Color online)
Analysis of the principal arc for the driving $f(t)=\cos(\Omega t)$ and
system-bath coupling with $\Delta=0.5\Omega$.
(a) Fourier transform of the interference pattern,
$W(\tau_\epsilon,\tau_A)$ along the principal arc $\tau_A =
2F(\tau_\epsilon/2)$ for $\sigma_x$ coupling and $\alpha=0.05$.
The symbols show numerical results for different temperatures $1/\beta$, while the
straight lines are fits to an exponential decay.
(b) Decay rate $\lambda$ for temperature $1/\beta=0.5\hbar\Omega$ as a function of the
dissipation strength $\alpha$. The error bars are determined by slightly
varying the fit range.
(c) Decay rate as a function of the temperature for dissipation strengths
$\alpha=0.05$.}
\label{fig:3}
\end{figure}
A promising application of LZSM interferometry is to determine microscopic
model parameters such as the dimensionless dissipation strength $\alpha$.
In Ref.~\cite{Forster2014a}, this was performed by comparing the decay of
the arc structure of measured LZSM patterns with corresponding theoretical
data.  This application raises interest in the corresponding decay rates as
a function of the bath parameters which are the dissipation strength and
the temperature.  Some examples for the arc decay are shown in
Fig.~\ref{fig:3}(a).  It can be appreciated that in the vicinity of
$\tau_\epsilon\approx T/4$, the Fourier amplitude decays exponentially,
$W(\tau_\epsilon) \propto \exp(-\lambda\tau_\epsilon)$, where $\lambda$ can
be determined by a numerical fit procedure.  Figures~\ref{fig:3}(b) and
\ref{fig:3}(c) show the result as a function of the dissipation strength
and the temperature, respectively. 

For longitudinal bath coupling, the rate exhibits a rather weak parameter
dependence.  A possible reason for this is that dissipative decays happen
mainly during the short stages when the levels cross.  Therefore the effective
decoherence rate is always much smaller than the ``natural'' width of
the asymmetric peaks given by $\Delta_n$, cf.\ Fig.~\ref{fig:P(e)}(a)
and Eq.~\eqref{Pz}.  At first sight, this weak parameter dependence seems
not in accordance with the LZSM patterns for open quantum dots with a bath
coupling via $\sigma_z$ \cite{Forster2014a}.  Notice however that the open double
quantum dot used there is beyond the present model.  First, the description
of electron transport requires one to take more states and different electron
numbers into account, especially when also spin effects play a role.
Second, there the dot-lead coupling is responsible for the main dissipative
effects, while the bath coupling represents a perturbation and does not
influence the qualitative behavior.

For the transverse bath coupling via $\sigma_x$, by contrast, $\lambda$
grows significantly and monotonically with the dissipation strength
$\alpha$, a feature that is essential for the fixing of
$\alpha$ from measured data.  The behavior as a function of the temperature
is more involved and even non-monotonic.  For very low temperatures, the
decay rate starts with a value $\lambda \approx 0.4\Omega$, followed by s
steep increase until the thermal energy matches the photon energy,
$k_BT\approx \hbar\Omega$.  Then a slow decay sets in which lasts until
eventually the range of exponential decay becomes so small that the fitting
procedure is no longer reasonable.

\section{Conclusions}

We have developed a comprehensive picture of LZSM interference
for the spin-boson model and thereby extended previous results to arbitrary
shapes of the periodic driving and a generalized qubit-bath coupling.  Our
central quantity of interest was the time-averaged population of the
excited state of the undriven qubit.  For its numerical computation, we
have employed a Bloch-Redfield master equation decomposed into the Floquet
states of the driven qubit, while avoiding any rotating-wave approximation
even in its moderate form.  Thus, our long-time solution contains the full
information about the coherences.

The interference patterns in ``real space'', i.e., as a function of the
detuning and the driving amplitude turned out to be governed by the qubit
operator that couples to the environmental degrees of freedom.  By
contrast, the shape of the driving is of minor relevance.  In particular,
we found that for a bath coupling that is transverse with respect to the
driving, the resonances are Lorentzians, while they possess an
anti-symmetric structure in the longitudinal case.  By a mapping to an
effective static Hamiltonian we have obtained Bloch equations which yield
expression for the LZSM patterns in agreement with numerical results.  As a
further feature, the LZSM pattern exhibits a triangular background which
can be explained within an adiabatic approximation for the full
time-dependent Bloch equations.  Moreover, in the presence of both a
transverse and a longitudinal bath coupling, the influence of the
transverse coupling prevails.

The Fourier transform of the LZSM patterns provide a complementary
picture.  While the bath coupling is of minor influence, the observed arc
structure reflects the shape of the driving, as can be predicted
from the solution of our effective Bloch equations.  For a driving with
time-reversal symmetry, the arcs are given by the integral of the driving.
In addition, they may develop side branches which can be explained within
our analytical approach, but their determination requires the moderate
effort of numerically solving a transcendental equation.  The same
numerical procedure also serves for the case of asymmetric driving.

A promising application of LZSM interferometry is the fixing of
dissipative parameters by comparing the arc decay for experimental and
theoretical data.  In this spirit, we have performed the corresponding
theoretical calculations.  They show that for transverse bath coupling, the
decay rate increases significantly with dissipation strength and
temperature, as long as the thermal energy does not exceed the energy
quantum of the driving.  Thus, in particular for predominantly transverse
coupling and low temperatures, LZSM interference represents a useful tools
for analyzing decoherence properties.  For purely longitudinal bath
coupling, by contrast, the arcs decay depends only weakly on dissipation.

Our investigation reveals that already the LZSM pattern of a qubit is quite
intriguing.  It may become even more involved for Landau-Zener scenarios
with three or more levels \cite{Kiselev2013, Satanin2012a} which are
relevant when spin effects enter \cite{Ribeiro2013a} or for a qubit
that couples to additional degrees of freedom such as, e.g., an exciton in
a photonic crystal with a coupling modulated by a surface-acoustic
\cite{Blattmann2014a}.  LZSM interferometry for such setups represents an
emerging field of investigation.

\begin{acknowledgments}
We like to thank Florian Forster and Stefan Ludwig for helpful discussions.
This work was supported by the Deutsche Forschungsgemeinschaft via
Sonderforschungsbereich SFB 631 (Project A5) and by the Spanish Ministry of
Economy and Competitiveness through grant no.\ MAT2011-24331.
\end{acknowledgments}

\appendix

\section{Bloch equations}
\label{app:Bloch}

In order to derive an equation of motion for the \textit{time-independent}
qubit, we start from the master equation \eqref{eq:master}. and notice that
for the Ohmic spectral density $J(\omega) = 2\pi\alpha\omega$, the
anti-symmetric bath correlation function \eqref{eq:antisym} becomes
$A(\tau) = 2\pi\alpha\delta'(\tau)$.  This has for the $\tau$-integral in
Eq.~\eqref{eq:master} the consequence that the Heisenberg operator $\tilde
X$ turns into its time-derivative evaluated at $\tau=0$.  Thus it can be
expressed by the commutator $i[H,X]$ and we obtain
\begin{equation}
\dot\rho = -i[H,\rho]
  -\frac{1}{4}[X,Q,\rho]] + \frac{\pi\alpha}{4} [X,\{[H,X],\rho\}] ,
\label{staticME}
\end{equation}
where the second term depends on the coherent qubit dynamics via the operator
\begin{equation}
Q = \frac{1}{2} \int_{-\infty}^{+\infty} d\tau\, S(\tau) \tilde X(-\tau) .
\label{Q}
\end{equation}
Since all analytical results of the main paper can be mapped by a
permutation of the Pauli matrices to a qubit in its eigenbasis with a
qubit-bath coupling via either $X=\sigma_x$ or $X=\sigma_z$, we consider
the Hamiltonian
\begin{equation}
  H = \frac{E}{2} \sigma_z .
\end{equation}

For $X=\sigma_x$, the Heisenberg operator in Eq.~\eqref{Q} reads
$\tilde\sigma_x(-\tau) = \sigma_x\cos(E\tau) - \sigma_y\sin(E\tau)$.  With
this expressions at hand, it is straightforward to evaluate the operator
$Q$ and to map the master equation \eqref{staticME} to an equation of
motion for the Bloch vector $\vec s = \tr(\vec\sigma\rho)$.  After some
algebra and a rotating-wave approximation, we find the Bloch equation
\begin{equation}
 \label{eq:Bloch}
 \frac{d}{dt} \vec{s}
= 
  \begin{pmatrix}
     -\Gamma/2 & E  & 0 \\
     -E & -\Gamma/2 & 0 \\
     0         & 0         & -\Gamma
  \end{pmatrix} \vec{s} 
+ \begin{pmatrix} 0 \\ 0 \\ \pi\alpha E \end{pmatrix},
\end{equation}
where the rate
\begin{align}
\label{app:Gamma}
\Gamma = \pi\alpha E \coth(\beta E/2)
\end{align}
depends on the qubit splitting and at low temperatures, $kT\ll E$, it
becomes $\Gamma = \pi\alpha|E|$.

For $\sigma_z$ coupling, the Heisenberg operator of the bath coupling is
time independent, $\tilde\sigma_z(-\tau) = \sigma_z$, so that the
$\tau$-integral yields the
Fourier transform of the symmetric spectral density at zero frequency.
Moreover, the last term of the master equation \eqref{staticME} vanishes.
Accordingly, the Bloch equation is homogeneous and reads
\begin{equation}
 \label{eq:BlochZ}
 \frac{d}{dt} \vec{s}
= 
  \begin{pmatrix}
     -\Gamma\varphi & E         & 0 \\
     -E        & -\Gamma\varphi & 0 \\
     0         & 0         & 0
  \end{pmatrix} \vec{s} ,
\end{equation}
where the dephasing rate
\begin{align}
\label{app:Gammaphi}
\Gamma_\varphi = 4\pi\alpha kT
\end{align}
vanishes in the zero-temperature limit.  Notice that the $z$-component of the
Bloch vector is conserved.

\bibliography{literature}

\begin{thebibliography}{30}%
\makeatletter
\providecommand \@ifxundefined [1]{%
 \@ifx{#1\undefined}
}%
\providecommand \@ifnum [1]{%
 \ifnum #1\expandafter \@firstoftwo
 \else \expandafter \@secondoftwo
 \fi
}%
\providecommand \@ifx [1]{%
 \ifx #1\expandafter \@firstoftwo
 \else \expandafter \@secondoftwo
 \fi
}%
\providecommand \natexlab [1]{#1}%
\providecommand \enquote  [1]{``#1''}%
\providecommand \bibnamefont  [1]{#1}%
\providecommand \bibfnamefont [1]{#1}%
\providecommand \citenamefont [1]{#1}%
\providecommand \href@noop [0]{\@secondoftwo}%
\providecommand \href [0]{\begingroup \@sanitize@url \@href}%
\providecommand \@href[1]{\@@startlink{#1}\@@href}%
\providecommand \@@href[1]{\endgroup#1\@@endlink}%
\providecommand \@sanitize@url [0]{\catcode `\\12\catcode `\$12\catcode
  `\&12\catcode `\#12\catcode `\^12\catcode `\_12\catcode `\%12\relax}%
\providecommand \@@startlink[1]{}%
\providecommand \@@endlink[0]{}%
\providecommand \url  [0]{\begingroup\@sanitize@url \@url }%
\providecommand \@url [1]{\endgroup\@href {#1}{\urlprefix }}%
\providecommand \urlprefix  [0]{URL }%
\providecommand \Eprint [0]{\href }%
\providecommand \doibase [0]{http://dx.doi.org/}%
\providecommand \selectlanguage [0]{\@gobble}%
\providecommand \bibinfo  [0]{\@secondoftwo}%
\providecommand \bibfield  [0]{\@secondoftwo}%
\providecommand \translation [1]{[#1]}%
\providecommand \BibitemOpen [0]{}%
\providecommand \bibitemStop [0]{}%
\providecommand \bibitemNoStop [0]{.\EOS\space}%
\providecommand \EOS [0]{\spacefactor3000\relax}%
\providecommand \BibitemShut  [1]{\csname bibitem#1\endcsname}%
\let\auto@bib@innerbib\@empty
\bibitem [{\citenamefont {Landau}(1932)}]{Landau1932a}%
  \BibitemOpen
  \bibfield  {author} {\bibinfo {author} {\bibfnamefont {L.~D.}\ \bibnamefont
  {Landau}},\ }\href@noop {} {\bibfield  {journal} {\bibinfo  {journal} {Phys.
  Z. Sowjetunion}\ }\textbf {\bibinfo {volume} {2}},\ \bibinfo {pages} {46}
  (\bibinfo {year} {1932})}\BibitemShut {NoStop}%
\bibitem [{\citenamefont {Zener}(1932)}]{Zener1932a}%
  \BibitemOpen
  \bibfield  {author} {\bibinfo {author} {\bibfnamefont {C.}~\bibnamefont
  {Zener}},\ }\href@noop {} {\bibfield  {journal} {\bibinfo  {journal} {Proc.
  R. Soc. London, Ser. A}\ }\textbf {\bibinfo {volume} {137}},\ \bibinfo
  {pages} {696} (\bibinfo {year} {1932})}\BibitemShut {NoStop}%
\bibitem [{\citenamefont {Stueckelberg}(1932)}]{Stueckelberg1932a}%
  \BibitemOpen
  \bibfield  {author} {\bibinfo {author} {\bibfnamefont {E.~C.~G.}\
  \bibnamefont {Stueckelberg}},\ }\href@noop {} {\bibfield  {journal} {\bibinfo
   {journal} {Helv. Phys. Acta}\ }\textbf {\bibinfo {volume} {5}},\ \bibinfo
  {pages} {369} (\bibinfo {year} {1932})}\BibitemShut {NoStop}%
\bibitem [{\citenamefont {Majorana}(1932)}]{Majorana1932a}%
  \BibitemOpen
  \bibfield  {author} {\bibinfo {author} {\bibfnamefont {E.}~\bibnamefont
  {Majorana}},\ }\href@noop {} {\bibfield  {journal} {\bibinfo  {journal}
  {Nuovo Cimento}\ }\textbf {\bibinfo {volume} {9}},\ \bibinfo {pages} {43}
  (\bibinfo {year} {1932})}\BibitemShut {NoStop}%
\bibitem [{\citenamefont {Shevchenko}\ \emph {et~al.}(2010)\citenamefont
  {Shevchenko}, \citenamefont {Ashhab},\ and\ \citenamefont
  {Nori}}]{Shevchenko2010a}%
  \BibitemOpen
  \bibfield  {author} {\bibinfo {author} {\bibfnamefont {S.~N.}\ \bibnamefont
  {Shevchenko}}, \bibinfo {author} {\bibfnamefont {S.}~\bibnamefont {Ashhab}},
  \ and\ \bibinfo {author} {\bibfnamefont {F.}~\bibnamefont {Nori}},\ }\href
  {\doibase 10.1016/j.physrep.2010.03.002} {\bibfield  {journal} {\bibinfo
  {journal} {Phys. Rep.}\ }\textbf {\bibinfo {volume} {492}},\ \bibinfo {pages}
  {1} (\bibinfo {year} {2010})}\BibitemShut {NoStop}%
\bibitem [{\citenamefont {Oliver}\ \emph {et~al.}(2005)\citenamefont {Oliver},
  \citenamefont {Yu}, \citenamefont {Lee}, \citenamefont {Berggren},
  \citenamefont {Levitov},\ and\ \citenamefont {Orlando}}]{Oliver2005a}%
  \BibitemOpen
  \bibfield  {author} {\bibinfo {author} {\bibfnamefont {W.~D.}\ \bibnamefont
  {Oliver}}, \bibinfo {author} {\bibfnamefont {Y.}~\bibnamefont {Yu}}, \bibinfo
  {author} {\bibfnamefont {J.~C.}\ \bibnamefont {Lee}}, \bibinfo {author}
  {\bibfnamefont {K.~K.}\ \bibnamefont {Berggren}}, \bibinfo {author}
  {\bibfnamefont {L.~S.}\ \bibnamefont {Levitov}}, \ and\ \bibinfo {author}
  {\bibfnamefont {T.~P.}\ \bibnamefont {Orlando}},\ }\href@noop {} {\bibfield
  {journal} {\bibinfo  {journal} {Science}\ }\textbf {\bibinfo {volume}
  {310}},\ \bibinfo {pages} {1653} (\bibinfo {year} {2005})}\BibitemShut
  {NoStop}%
\bibitem [{\citenamefont {Sillanp\"a\"a}\ \emph {et~al.}(2006)\citenamefont
  {Sillanp\"a\"a}, \citenamefont {Lehtinen}, \citenamefont {Paila},
  \citenamefont {Makhlin},\ and\ \citenamefont {Hakonen}}]{Sillanpaa2006a}%
  \BibitemOpen
  \bibfield  {author} {\bibinfo {author} {\bibfnamefont {M.}~\bibnamefont
  {Sillanp\"a\"a}}, \bibinfo {author} {\bibfnamefont {T.}~\bibnamefont
  {Lehtinen}}, \bibinfo {author} {\bibfnamefont {A.}~\bibnamefont {Paila}},
  \bibinfo {author} {\bibfnamefont {Y.}~\bibnamefont {Makhlin}}, \ and\
  \bibinfo {author} {\bibfnamefont {P.}~\bibnamefont {Hakonen}},\ }\href@noop
  {} {\bibfield  {journal} {\bibinfo  {journal} {Phys. Rev. Lett.}\ }\textbf
  {\bibinfo {volume} {96}},\ \bibinfo {pages} {187002} (\bibinfo {year}
  {2006})}\BibitemShut {NoStop}%
\bibitem [{\citenamefont {Wilson}\ \emph {et~al.}(2007)\citenamefont {Wilson},
  \citenamefont {Duty}, \citenamefont {Persson}, \citenamefont {Sandberg},
  \citenamefont {Johansson},\ and\ \citenamefont {Delsing}}]{Wilson2007b}%
  \BibitemOpen
  \bibfield  {author} {\bibinfo {author} {\bibfnamefont {C.~M.}\ \bibnamefont
  {Wilson}}, \bibinfo {author} {\bibfnamefont {T.}~\bibnamefont {Duty}},
  \bibinfo {author} {\bibfnamefont {F.}~\bibnamefont {Persson}}, \bibinfo
  {author} {\bibfnamefont {M.}~\bibnamefont {Sandberg}}, \bibinfo {author}
  {\bibfnamefont {G.}~\bibnamefont {Johansson}}, \ and\ \bibinfo {author}
  {\bibfnamefont {P.}~\bibnamefont {Delsing}},\ }\href@noop {} {\bibfield
  {journal} {\bibinfo  {journal} {Phys. Rev. Lett.}\ }\textbf {\bibinfo
  {volume} {98}},\ \bibinfo {pages} {257003} (\bibinfo {year}
  {2007})}\BibitemShut {NoStop}%
\bibitem [{\citenamefont {Berns}\ \emph {et~al.}(2008)\citenamefont {Berns},
  \citenamefont {Rudner}, \citenamefont {Valenzuela}, \citenamefont {Berggren},
  \citenamefont {Oliver}, \citenamefont {Levitov},\ and\ \citenamefont
  {Orlando}}]{Berns2008a}%
  \BibitemOpen
  \bibfield  {author} {\bibinfo {author} {\bibfnamefont {D.~M.}\ \bibnamefont
  {Berns}}, \bibinfo {author} {\bibfnamefont {M.~S.}\ \bibnamefont {Rudner}},
  \bibinfo {author} {\bibfnamefont {S.~O.}\ \bibnamefont {Valenzuela}},
  \bibinfo {author} {\bibfnamefont {K.~K.}\ \bibnamefont {Berggren}}, \bibinfo
  {author} {\bibfnamefont {W.~D.}\ \bibnamefont {Oliver}}, \bibinfo {author}
  {\bibfnamefont {L.~S.}\ \bibnamefont {Levitov}}, \ and\ \bibinfo {author}
  {\bibfnamefont {T.~P.}\ \bibnamefont {Orlando}},\ }\href@noop {} {\bibfield
  {journal} {\bibinfo  {journal} {Nature (London)}\ }\textbf {\bibinfo {volume}
  {455}},\ \bibinfo {pages} {51} (\bibinfo {year} {2008})}\BibitemShut
  {NoStop}%
\bibitem [{\citenamefont {Stehlik}\ \emph {et~al.}(2012)\citenamefont
  {Stehlik}, \citenamefont {Dovzhenko}, \citenamefont {Petta}, \citenamefont
  {Johansson}, \citenamefont {Nori}, \citenamefont {Lu},\ and\ \citenamefont
  {Gossard}}]{Stehlik2012a}%
  \BibitemOpen
  \bibfield  {author} {\bibinfo {author} {\bibfnamefont {J.}~\bibnamefont
  {Stehlik}}, \bibinfo {author} {\bibfnamefont {Y.}~\bibnamefont {Dovzhenko}},
  \bibinfo {author} {\bibfnamefont {J.~R.}\ \bibnamefont {Petta}}, \bibinfo
  {author} {\bibfnamefont {J.~R.}\ \bibnamefont {Johansson}}, \bibinfo {author}
  {\bibfnamefont {F.}~\bibnamefont {Nori}}, \bibinfo {author} {\bibfnamefont
  {H.}~\bibnamefont {Lu}}, \ and\ \bibinfo {author} {\bibfnamefont {A.~C.}\
  \bibnamefont {Gossard}},\ }\href@noop {} {\bibfield  {journal} {\bibinfo
  {journal} {Phys. Rev. B}\ }\textbf {\bibinfo {volume} {86}},\ \bibinfo
  {pages} {121303(R)} (\bibinfo {year} {2012})}\BibitemShut {NoStop}%
\bibitem [{\citenamefont {Dupont-Ferrier}\ \emph {et~al.}(2013)\citenamefont
  {Dupont-Ferrier}, \citenamefont {Roche}, \citenamefont {Voisin},
  \citenamefont {Jehl}, \citenamefont {Wacquez}, \citenamefont {Vinet},
  \citenamefont {Sanquer},\ and\ \citenamefont
  {De~Franceschi}}]{Dupont-Ferrier2013}%
  \BibitemOpen
  \bibfield  {author} {\bibinfo {author} {\bibfnamefont {E.}~\bibnamefont
  {Dupont-Ferrier}}, \bibinfo {author} {\bibfnamefont {B.}~\bibnamefont
  {Roche}}, \bibinfo {author} {\bibfnamefont {B.}~\bibnamefont {Voisin}},
  \bibinfo {author} {\bibfnamefont {X.}~\bibnamefont {Jehl}}, \bibinfo {author}
  {\bibfnamefont {R.}~\bibnamefont {Wacquez}}, \bibinfo {author} {\bibfnamefont
  {M.}~\bibnamefont {Vinet}}, \bibinfo {author} {\bibfnamefont
  {M.}~\bibnamefont {Sanquer}}, \ and\ \bibinfo {author} {\bibfnamefont
  {S.}~\bibnamefont {De~Franceschi}},\ }\href {\doibase
  10.1103/PhysRevLett.110.136802} {\bibfield  {journal} {\bibinfo  {journal}
  {Phys. Rev. Lett.}\ }\textbf {\bibinfo {volume} {110}},\ \bibinfo {pages}
  {136802} (\bibinfo {year} {2013})}\BibitemShut {NoStop}%
\bibitem [{\citenamefont {Li}\ \emph {et~al.}(2013)\citenamefont {Li},
  \citenamefont {Silveri}, \citenamefont {Kumar}, \citenamefont {Pirkkalainen},
  \citenamefont {Veps\"al\"ainen}, \citenamefont {Chien}, \citenamefont
  {Tuorila}, \citenamefont {Sillanp\"a\"a}, \citenamefont {Hakonen},
  \citenamefont {Thuneberg},\ and\ \citenamefont {Paraoanu}}]{Li2013a}%
  \BibitemOpen
  \bibfield  {author} {\bibinfo {author} {\bibfnamefont {J.}~\bibnamefont
  {Li}}, \bibinfo {author} {\bibfnamefont {M.~P.}\ \bibnamefont {Silveri}},
  \bibinfo {author} {\bibfnamefont {K.~S.}\ \bibnamefont {Kumar}}, \bibinfo
  {author} {\bibfnamefont {J.-M.}\ \bibnamefont {Pirkkalainen}}, \bibinfo
  {author} {\bibfnamefont {A.}~\bibnamefont {Veps\"al\"ainen}}, \bibinfo
  {author} {\bibfnamefont {W.~C.}\ \bibnamefont {Chien}}, \bibinfo {author}
  {\bibfnamefont {J.}~\bibnamefont {Tuorila}}, \bibinfo {author} {\bibfnamefont
  {M.~A.}\ \bibnamefont {Sillanp\"a\"a}}, \bibinfo {author} {\bibfnamefont
  {P.~J.}\ \bibnamefont {Hakonen}}, \bibinfo {author} {\bibfnamefont {E.~V.}\
  \bibnamefont {Thuneberg}}, \ and\ \bibinfo {author} {\bibfnamefont {G.~S.}\
  \bibnamefont {Paraoanu}},\ }\href@noop {} {\bibfield  {journal} {\bibinfo
  {journal} {Nature Comm.}\ }\textbf {\bibinfo {volume} {4}},\ \bibinfo {pages}
  {1420} (\bibinfo {year} {2013})}\BibitemShut {NoStop}%
\bibitem [{\citenamefont {Forster}\ \emph {et~al.}(2014)\citenamefont
  {Forster}, \citenamefont {Petersen}, \citenamefont {Manus}, \citenamefont
  {H{\"a}nggi}, \citenamefont {Schuh}, \citenamefont {Wegscheider},
  \citenamefont {Kohler},\ and\ \citenamefont {Ludwig}}]{Forster2014a}%
  \BibitemOpen
  \bibfield  {author} {\bibinfo {author} {\bibfnamefont {F.}~\bibnamefont
  {Forster}}, \bibinfo {author} {\bibfnamefont {G.}~\bibnamefont {Petersen}},
  \bibinfo {author} {\bibfnamefont {S.}~\bibnamefont {Manus}}, \bibinfo
  {author} {\bibfnamefont {P.}~\bibnamefont {H{\"a}nggi}}, \bibinfo {author}
  {\bibfnamefont {D.}~\bibnamefont {Schuh}}, \bibinfo {author} {\bibfnamefont
  {W.}~\bibnamefont {Wegscheider}}, \bibinfo {author} {\bibfnamefont
  {S.}~\bibnamefont {Kohler}}, \ and\ \bibinfo {author} {\bibfnamefont
  {S.}~\bibnamefont {Ludwig}},\ }\href@noop {} {\bibfield  {journal} {\bibinfo
  {journal} {Phys. Rev. Lett.}\ }\textbf {\bibinfo {volume} {112}},\ \bibinfo
  {pages} {116803} (\bibinfo {year} {2014})}\BibitemShut {NoStop}%
\bibitem [{\citenamefont {Rudner}\ \emph {et~al.}(2008)\citenamefont {Rudner},
  \citenamefont {Shytov}, \citenamefont {Levitov}, \citenamefont {Berns},
  \citenamefont {Oliver}, \citenamefont {Valenzuela},\ and\ \citenamefont
  {Orlando}}]{Rudner2008a}%
  \BibitemOpen
  \bibfield  {author} {\bibinfo {author} {\bibfnamefont {M.~S.}\ \bibnamefont
  {Rudner}}, \bibinfo {author} {\bibfnamefont {A.~V.}\ \bibnamefont {Shytov}},
  \bibinfo {author} {\bibfnamefont {L.~S.}\ \bibnamefont {Levitov}}, \bibinfo
  {author} {\bibfnamefont {D.~M.}\ \bibnamefont {Berns}}, \bibinfo {author}
  {\bibfnamefont {W.~D.}\ \bibnamefont {Oliver}}, \bibinfo {author}
  {\bibfnamefont {S.~O.}\ \bibnamefont {Valenzuela}}, \ and\ \bibinfo {author}
  {\bibfnamefont {T.~P.}\ \bibnamefont {Orlando}},\ }\href@noop {} {\bibfield
  {journal} {\bibinfo  {journal} {Phys. Rev. Lett.}\ }\textbf {\bibinfo
  {volume} {101}},\ \bibinfo {pages} {190502} (\bibinfo {year}
  {2008})}\BibitemShut {NoStop}%
\bibitem [{\citenamefont {Grifoni}\ and\ \citenamefont
  {H\"anggi}(1998)}]{Grifoni1998a}%
  \BibitemOpen
  \bibfield  {author} {\bibinfo {author} {\bibfnamefont {M.}~\bibnamefont
  {Grifoni}}\ and\ \bibinfo {author} {\bibfnamefont {P.}~\bibnamefont
  {H\"anggi}},\ }\href {\doibase 10.1016/S0370-1573(98)00022-2} {\bibfield
  {journal} {\bibinfo  {journal} {Phys. Rep.}\ }\textbf {\bibinfo {volume}
  {304}},\ \bibinfo {pages} {229} (\bibinfo {year} {1998})}\BibitemShut
  {NoStop}%
\bibitem [{\citenamefont {Kohler}\ \emph {et~al.}(1997)\citenamefont {Kohler},
  \citenamefont {Dittrich},\ and\ \citenamefont {H\"anggi}}]{Kohler1997a}%
  \BibitemOpen
  \bibfield  {author} {\bibinfo {author} {\bibfnamefont {S.}~\bibnamefont
  {Kohler}}, \bibinfo {author} {\bibfnamefont {T.}~\bibnamefont {Dittrich}}, \
  and\ \bibinfo {author} {\bibfnamefont {P.}~\bibnamefont {H\"anggi}},\
  }\href@noop {} {\bibfield  {journal} {\bibinfo  {journal} {Phys. Rev. E}\
  }\textbf {\bibinfo {volume} {55}},\ \bibinfo {pages} {300} (\bibinfo {year}
  {1997})}\BibitemShut {NoStop}%
\bibitem [{\citenamefont {Bl\"umel}\ \emph {et~al.}(1991)\citenamefont
  {Bl\"umel}, \citenamefont {Buchleitner}, \citenamefont {Graham},
  \citenamefont {Sirko}, \citenamefont {Smilansky},\ and\ \citenamefont
  {Walter}}]{Blumel1991a}%
  \BibitemOpen
  \bibfield  {author} {\bibinfo {author} {\bibfnamefont {R.}~\bibnamefont
  {Bl\"umel}}, \bibinfo {author} {\bibfnamefont {A.}~\bibnamefont
  {Buchleitner}}, \bibinfo {author} {\bibfnamefont {R.}~\bibnamefont {Graham}},
  \bibinfo {author} {\bibfnamefont {L.}~\bibnamefont {Sirko}}, \bibinfo
  {author} {\bibfnamefont {U.}~\bibnamefont {Smilansky}}, \ and\ \bibinfo
  {author} {\bibfnamefont {H.}~\bibnamefont {Walter}},\ }\href@noop {}
  {\bibfield  {journal} {\bibinfo  {journal} {Phys. Rev. A}\ }\textbf {\bibinfo
  {volume} {44}},\ \bibinfo {pages} {4521} (\bibinfo {year}
  {1991})}\BibitemShut {NoStop}%
\bibitem [{\citenamefont {Lehmann}\ \emph {et~al.}(2002)\citenamefont
  {Lehmann}, \citenamefont {Kohler}, \citenamefont {H\"anggi},\ and\
  \citenamefont {Nitzan}}]{Lehmann2002b}%
  \BibitemOpen
  \bibfield  {author} {\bibinfo {author} {\bibfnamefont {J.}~\bibnamefont
  {Lehmann}}, \bibinfo {author} {\bibfnamefont {S.}~\bibnamefont {Kohler}},
  \bibinfo {author} {\bibfnamefont {P.}~\bibnamefont {H\"anggi}}, \ and\
  \bibinfo {author} {\bibfnamefont {A.}~\bibnamefont {Nitzan}},\ }\href@noop {}
  {\bibfield  {journal} {\bibinfo  {journal} {Phys. Rev. Lett.}\ }\textbf
  {\bibinfo {volume} {88}},\ \bibinfo {pages} {228305} (\bibinfo {year}
  {2002})}\BibitemShut {NoStop}%
\bibitem [{\citenamefont {Gro{\ss}mann}\ and\ \citenamefont
  {H\"anggi}(1992)}]{Grossmann1992a}%
  \BibitemOpen
  \bibfield  {author} {\bibinfo {author} {\bibfnamefont {F.}~\bibnamefont
  {Gro{\ss}mann}}\ and\ \bibinfo {author} {\bibfnamefont {P.}~\bibnamefont
  {H\"anggi}},\ }\href@noop {} {\bibfield  {journal} {\bibinfo  {journal}
  {Europhys. Lett.}\ }\textbf {\bibinfo {volume} {18}},\ \bibinfo {pages} {571}
  (\bibinfo {year} {1992})}\BibitemShut {NoStop}%
\bibitem [{\citenamefont {Kayanuma}(1994)}]{Kayanuma1994a}%
  \BibitemOpen
  \bibfield  {author} {\bibinfo {author} {\bibfnamefont {Y.}~\bibnamefont
  {Kayanuma}},\ }\href@noop {} {\bibfield  {journal} {\bibinfo  {journal}
  {Phys. Rev. A}\ }\textbf {\bibinfo {volume} {50}},\ \bibinfo {pages} {843}
  (\bibinfo {year} {1994})}\BibitemShut {NoStop}%
\bibitem [{\citenamefont {Satanin}\ \emph {et~al.}()\citenamefont {Satanin},
  \citenamefont {Denisenko}, \citenamefont {Gelman},\ and\ \citenamefont
  {Nori}}]{Satanin2014a}%
  \BibitemOpen
  \bibfield  {author} {\bibinfo {author} {\bibfnamefont {A.~M.}\ \bibnamefont
  {Satanin}}, \bibinfo {author} {\bibfnamefont {M.~V.}\ \bibnamefont
  {Denisenko}}, \bibinfo {author} {\bibfnamefont {A.~I.}\ \bibnamefont
  {Gelman}}, \ and\ \bibinfo {author} {\bibfnamefont {F.}~\bibnamefont
  {Nori}},\ }\href@noop {} {}\Eprint {http://arxiv.org/abs/arXiv:1305.4800
  [cond-mat]} {arXiv:1305.4800 [cond-mat]} \BibitemShut {NoStop}%
\bibitem [{\citenamefont {Strass}\ \emph {et~al.}(2005)\citenamefont {Strass},
  \citenamefont {H\"anggi},\ and\ \citenamefont {Kohler}}]{Strass2005b}%
  \BibitemOpen
  \bibfield  {author} {\bibinfo {author} {\bibfnamefont {M.}~\bibnamefont
  {Strass}}, \bibinfo {author} {\bibfnamefont {P.}~\bibnamefont {H\"anggi}}, \
  and\ \bibinfo {author} {\bibfnamefont {S.}~\bibnamefont {Kohler}},\
  }\href@noop {} {\bibfield  {journal} {\bibinfo  {journal} {Phys. Rev. Lett.}\
  }\textbf {\bibinfo {volume} {95}},\ \bibinfo {pages} {130601} (\bibinfo
  {year} {2005})}\BibitemShut {NoStop}%
\bibitem [{\citenamefont {Izmalkov}\ \emph {et~al.}(2008)\citenamefont
  {Izmalkov}, \citenamefont {van~der Ploeg}, \citenamefont {Shevchenko},
  \citenamefont {Grajcar}, \citenamefont {Il'ichev}, \citenamefont {H\"ubner},
  \citenamefont {Omelyanchouk},\ and\ \citenamefont {Meyer}}]{Izmalkov2008a}%
  \BibitemOpen
  \bibfield  {author} {\bibinfo {author} {\bibfnamefont {A.}~\bibnamefont
  {Izmalkov}}, \bibinfo {author} {\bibfnamefont {S.~H.~W.}\ \bibnamefont
  {van~der Ploeg}}, \bibinfo {author} {\bibfnamefont {S.~N.}\ \bibnamefont
  {Shevchenko}}, \bibinfo {author} {\bibfnamefont {M.}~\bibnamefont {Grajcar}},
  \bibinfo {author} {\bibfnamefont {E.}~\bibnamefont {Il'ichev}}, \bibinfo
  {author} {\bibfnamefont {U.}~\bibnamefont {H\"ubner}}, \bibinfo {author}
  {\bibfnamefont {A.~N.}\ \bibnamefont {Omelyanchouk}}, \ and\ \bibinfo
  {author} {\bibfnamefont {H.-G.}\ \bibnamefont {Meyer}},\ }\href@noop {}
  {\bibfield  {journal} {\bibinfo  {journal} {Phys. Rev. Lett.}\ }\textbf
  {\bibinfo {volume} {101}},\ \bibinfo {pages} {017003} (\bibinfo {year}
  {2008})}\BibitemShut {NoStop}%
\bibitem [{\citenamefont {Thingna}\ \emph {et~al.}(2012)\citenamefont
  {Thingna}, \citenamefont {Wang},\ and\ \citenamefont
  {H\"anggi}}]{Thingna2012a}%
  \BibitemOpen
  \bibfield  {author} {\bibinfo {author} {\bibfnamefont {J.}~\bibnamefont
  {Thingna}}, \bibinfo {author} {\bibfnamefont {J.-S.}\ \bibnamefont {Wang}}, \
  and\ \bibinfo {author} {\bibfnamefont {P.}~\bibnamefont {H\"anggi}},\
  }\href@noop {} {\bibfield  {journal} {\bibinfo  {journal} {J. Chem. Phys.}\
  }\textbf {\bibinfo {volume} {136}},\ \bibinfo {pages} {194110} (\bibinfo
  {year} {2012})}\BibitemShut {NoStop}%
\bibitem [{\citenamefont {Fonseca-Romero}\ \emph {et~al.}(2004)\citenamefont
  {Fonseca-Romero}, \citenamefont {Kohler},\ and\ \citenamefont
  {H\"anggi}}]{Fonseca2004a}%
  \BibitemOpen
  \bibfield  {author} {\bibinfo {author} {\bibfnamefont {K.~M.}\ \bibnamefont
  {Fonseca-Romero}}, \bibinfo {author} {\bibfnamefont {S.}~\bibnamefont
  {Kohler}}, \ and\ \bibinfo {author} {\bibfnamefont {P.}~\bibnamefont
  {H\"anggi}},\ }\href@noop {} {\bibfield  {journal} {\bibinfo  {journal}
  {Chem. Phys.}\ }\textbf {\bibinfo {volume} {296}},\ \bibinfo {pages} {307}
  (\bibinfo {year} {2004})}\BibitemShut {NoStop}%
\bibitem [{\citenamefont {Stace}\ \emph {et~al.}(2005)\citenamefont {Stace},
  \citenamefont {Doherty},\ and\ \citenamefont {Barrett}}]{Stace2005}%
  \BibitemOpen
  \bibfield  {author} {\bibinfo {author} {\bibfnamefont {T.~M.}\ \bibnamefont
  {Stace}}, \bibinfo {author} {\bibfnamefont {A.~C.}\ \bibnamefont {Doherty}},
  \ and\ \bibinfo {author} {\bibfnamefont {S.~D.}\ \bibnamefont {Barrett}},\
  }\href@noop {} {\bibfield  {journal} {\bibinfo  {journal} {Phys. Rev. Lett.}\
  }\textbf {\bibinfo {volume} {95}},\ \bibinfo {pages} {106801} (\bibinfo
  {year} {2005})}\BibitemShut {NoStop}%
\bibitem [{\citenamefont {Kiselev}\ \emph {et~al.}(2013)\citenamefont
  {Kiselev}, \citenamefont {Kikoin},\ and\ \citenamefont
  {Kenmoe}}]{Kiselev2013}%
  \BibitemOpen
  \bibfield  {author} {\bibinfo {author} {\bibfnamefont {M.~N.}\ \bibnamefont
  {Kiselev}}, \bibinfo {author} {\bibfnamefont {K.}~\bibnamefont {Kikoin}}, \
  and\ \bibinfo {author} {\bibfnamefont {M.~B.}\ \bibnamefont {Kenmoe}},\
  }\href {\doibase 10.1209/0295-5075/104/57004} {\bibfield  {journal} {\bibinfo
   {journal} {EPL}\ }\textbf {\bibinfo {volume} {104}},\ \bibinfo {pages}
  {57004} (\bibinfo {year} {2013})}\BibitemShut {NoStop}%
\bibitem [{\citenamefont {Satanin}\ \emph {et~al.}(2012)\citenamefont
  {Satanin}, \citenamefont {Denisenko}, \citenamefont {Ashhab},\ and\
  \citenamefont {Nori}}]{Satanin2012a}%
  \BibitemOpen
  \bibfield  {author} {\bibinfo {author} {\bibfnamefont {A.~M.}\ \bibnamefont
  {Satanin}}, \bibinfo {author} {\bibfnamefont {M.~V.}\ \bibnamefont
  {Denisenko}}, \bibinfo {author} {\bibfnamefont {S.}~\bibnamefont {Ashhab}}, \
  and\ \bibinfo {author} {\bibfnamefont {F.}~\bibnamefont {Nori}},\ }\href@noop
  {} {\bibfield  {journal} {\bibinfo  {journal} {Phys. Rev. B}\ }\textbf
  {\bibinfo {volume} {85}},\ \bibinfo {pages} {184524} (\bibinfo {year}
  {2012})}\BibitemShut {NoStop}%
\bibitem [{\citenamefont {Ribeiro}\ \emph {et~al.}(2013)\citenamefont
  {Ribeiro}, \citenamefont {Burkard}, \citenamefont {Petta}, \citenamefont
  {Lu},\ and\ \citenamefont {Gossard}}]{Ribeiro2013a}%
  \BibitemOpen
  \bibfield  {author} {\bibinfo {author} {\bibfnamefont {H.}~\bibnamefont
  {Ribeiro}}, \bibinfo {author} {\bibfnamefont {G.}~\bibnamefont {Burkard}},
  \bibinfo {author} {\bibfnamefont {J.~R.}\ \bibnamefont {Petta}}, \bibinfo
  {author} {\bibfnamefont {H.}~\bibnamefont {Lu}}, \ and\ \bibinfo {author}
  {\bibfnamefont {A.~C.}\ \bibnamefont {Gossard}},\ }\href@noop {} {\bibfield
  {journal} {\bibinfo  {journal} {Phys. Rev. Lett.}\ }\textbf {\bibinfo
  {volume} {110}},\ \bibinfo {pages} {086804} (\bibinfo {year}
  {2013})}\BibitemShut {NoStop}%
\bibitem [{\citenamefont {Blattmann}\ \emph {et~al.}(2014)\citenamefont
  {Blattmann}, \citenamefont {Krenner}, \citenamefont {Kohler},\ and\
  \citenamefont {H\"anggi}}]{Blattmann2014a}%
  \BibitemOpen
  \bibfield  {author} {\bibinfo {author} {\bibfnamefont {R.}~\bibnamefont
  {Blattmann}}, \bibinfo {author} {\bibfnamefont {H.~J.}\ \bibnamefont
  {Krenner}}, \bibinfo {author} {\bibfnamefont {S.}~\bibnamefont {Kohler}}, \
  and\ \bibinfo {author} {\bibfnamefont {P.}~\bibnamefont {H\"anggi}},\
  }\href@noop {} {\bibfield  {journal} {\bibinfo  {journal} {Phys. Rev. A}\
  }\textbf {\bibinfo {volume} {89}},\ \bibinfo {pages} {012327} (\bibinfo
  {year} {2014})}\BibitemShut {NoStop}%
\end{thebibliography}%

\end{document}